# New results on muon inelastic cross section and energy loss in rock

D.A. Timashkov, A.A. Petrukhin

*Moscow Engineering Physics Institute, Moscow, 115409, Russia*
(DATimashkov@ mephi.ru)

Various uncertainties in calculations of inelastic cross section and energy loss are considered. It is shown that widely used kinematic neglects and approximations result in deviations in calculations of these values. The obtained corrections increase with lepton mass, therefore possible consequences for tau-lepton are discussed, too.

## 1. Introduction

Inelastic scattering with nuclei plays important role in cosmic ray muon interaction and propagation at large depths of rock or water. In comparison with other electromagnetic processes inelastic cross section more slowly decreases with the increase of transfer momentum, and relative inelastic energy loss logarithmically rises with lepton energy. But in spite of abundance of various theoretical models and numerical calculations, the problem of correct estimation of contribution of inelastic processes to energy loss has not been finally solved (see reviews [1–3]). There are several main uncertainties which can influence the results of inelastic cross section and energy loss: 1) Absence of theory of inelastic interaction which describes inelastic cross section in the whole kinematic region as in perturbative so non-perturbative parts. 2) The use of approximate formulas for kinematic boundaries. 3) The neglect of lepton mass in cross section formulas. 4) Disregard of any nucleon resonance effects. 5) Calculation of nuclear effects in inelastic scattering with a simple correction function $A_{eff} = A^\alpha$.

Usually it is supposed that the main source of uncertainties in inelastic scattering is caused by the first reason. The recent results of description of inelastic lepton-proton cross section with taking into account of limiting dependences allow to essentially decrease such uncertainties [4]. In the present paper the attention is focused to remaining sources of probable corrections. In particular, it is shown that for such integral characteristic as total inelastic cross section and energy loss coefficient $b_{in}$ the choice of structure function model does not play a crucial role (if the model takes into account the main features of inelastic interaction) while the incorrect application of kinematic relations and the neglect of other remarkable effects can lead to significant distortions in the results even for ultra high energies of leptons.

## 2. Some kinematic aspects

Let us consider the kinematic boundaries for inelastic $lp$-scattering. The lower limit of transferred energy $\nu$ is derived from single pion production while the upper limit is obtained from the demand of stopping lepton after scattering:

$$m_\pi + \frac{m_\pi^2}{2M} \approx 144.7 \text{ MeV} \leq \nu \leq E - m_l = \nu_{max}, \qquad (1)$$

here $M$ is proton mass, $E$ is initial lepton energy. Traditionally (see [1–3]) the kinematic limits of squared transferred 4-momentum in inelastic scattering $Q^2$ are defined in a following way:

$$\frac{m_l^2 \nu^2}{EE'} < Q^2 < 2M\nu. \qquad (2)$$

Here $E'$ is a scattered lepton energy. These constraints are approximate and are obtained for ultra-relativistic leptons as before so *after* the scattering. The exact expression for lower limit of $Q^2$ has a following form:

$$Q^2_{min} = \frac{2m_l^2 \nu^2}{EE' + |\vec{p}||\vec{p}'| - m_l^2}, \qquad (3)$$

where $\vec{p}$, $\vec{p}'$ are lepton momentum before and after the scattering. As to the upper limit there are two conditions for $Q^2$: from kinematics of hadron vertex (see (2)) and from kinematics of lepton vertex:

$$Q^2 < 2EE' + 2|\vec{p}|\cdot|\vec{p}'| - 2m_l^2. \qquad (4)$$

On the basis of comparison of Eqs. (2) and (4), the kinematic range for transferred energy can be divided into two regions. In one of them (at $\nu < \nu_{tr}$) the maximum $Q^2$ is defined by Eq. (2). In another one, at $\nu > \nu_{tr}$, the condition of Eq. (4) works. The value of transition energy point coincides with maximum energy transfer for elastic scattering on nucleon:

$$\nu_{tr} = \vec{p}^2 \big/ \left( E + (M^2 + m_l^2)/(2M) \right). \qquad (5)$$

The use of the approximate expression (2) for any transferred energies leads to the extension of kinematic region beyond the permitted by conservation laws (see Fig. 1).

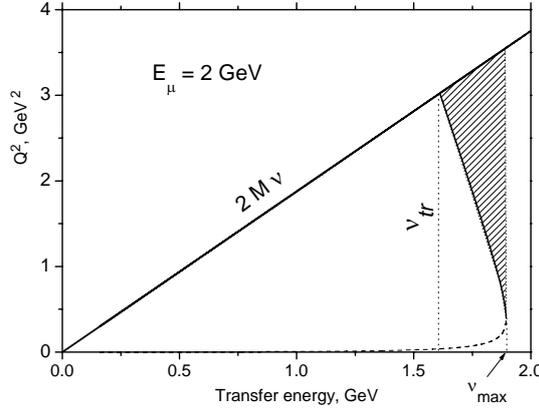

**Figure 1.** Kinematic regions for inelastic μp-scattering. Shaded area is a region forbidden by conservation laws.

Differential inelastic cross section and nucleon structure functions depend on two kinematic variables: transferred energy $\nu$ and squared 4-momentum $Q^2$:

$$\frac{d\sigma}{d\nu dQ^2} = \frac{2\pi\alpha^2}{\nu Q^4}\{Y_+ F_2 - Y_L F_L\}, \qquad (6)$$

where

$$Y_+ = 1 + (1-y)^2 + \frac{Mx_B y}{E} - 2y^2 \frac{m_l^2}{Q^2}\left(1 + \frac{Q^2}{\nu^2}\right), \quad Y_L = y^2\left(1 - \frac{2m_l^2}{Q^2}\right), \quad y = \nu/E, \qquad (7)$$

and $x_B = Q^2/(2M\nu)$ is the Bjorken variable.

Usually the lepton mass in Eq. (7) is neglected for high energy leptons. But since lower $Q^2$-limit (3) for high energies may be less than $2m_l^2$, the kinematic coefficient $Y_L$ can change its sign. Moreover, since the least values of $Q^2$ are reached at small values of $y$, in this region functions $Y_+$ and $Y_L$ have the following limits:

$$Y_+ \approx 6/\gamma^2; \quad Y_L \approx -2, \qquad (8)$$

and the differential cross section can be written in the following form:

$$\frac{d^2\sigma}{d\nu dQ^2} = \frac{4\pi\alpha_e^2}{\nu Q^4}\left\{F_L + \frac{3}{\gamma^2}F_2\right\}, \qquad (9)$$

where γ is a lepton Lorentz factor. Thus, for ultra-relativistic leptons the longitudinal constituent of the cross section gives the main contribution. Since $F_L/F_2 \sim Q^2/\nu^2$ (see [4, 5]), the differential cross section at low $Q^2$ becomes smaller than it follows from conventional relations neglecting lepton mass.

## 3. Results of calculations

For calculations of cross sections and energy loss the structure functions from paper [4] were used. This approach takes into account the dependence of inelastic form factors in three limiting cases: quasielastic scattering, photonuclear and low-$x_B$ limits. The resonances are taken into account in photonuclear cross section by means of the method described in [6]. Nuclear corrections are taken from a fit of NMC data [7] on muon-nuclei scattering. In Fig. 2 the results of calculations of the inelastic cross section $\sigma_{lA}$ and coefficient of energy loss $b_{in}$ in standard rock for three lepton generations using exact kinematic boundaries and relations are shown.

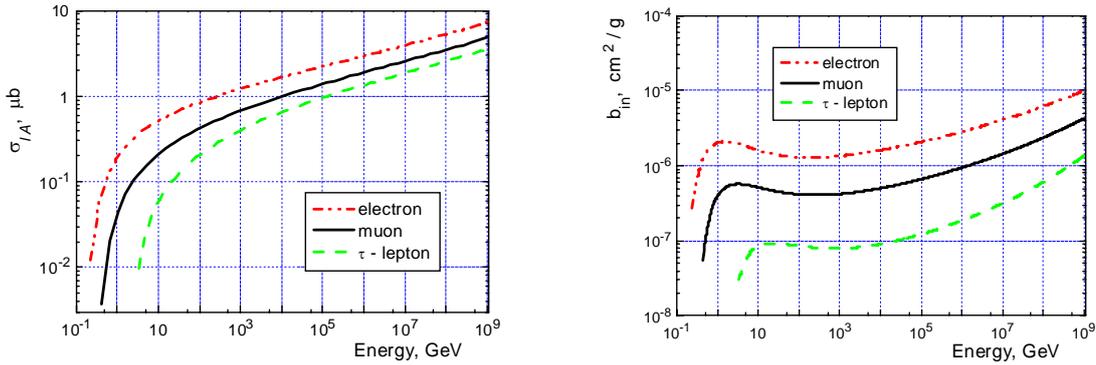

**Figure 2.** Inelastic cross section (left) and energy loss coefficient (right) for standard rock.

From the right figure one can note remarkable relations, which connect energy loss for different lepton types. In a broad range of lepton energies from 10 GeV to 10 PeV:

$$b_e \approx 3b_\mu, \qquad b_\mu \approx 5b_\tau. \qquad (10)$$

For investigation of the influence of various factors listed in the introduction, several schemes of energy loss coefficient calculation have been used: in the first one (simple), the items 2–5 (see Section 1) are taken into account in a conventional way (with approximate kinematic relations, without resonances and with a simple formula for nuclear corrections). In the another scheme (exact) the all corrections issued from items 2–5 are used. Schemes with number 2–5 represent the exact one without allowance the effect from corresponding item. Results for ratios of schemes 2–5 to exact one for muon and τ-lepton are presented in Fig.3.

It is remarkable that even for ultra-relativistic muon energies the correction to energy loss is quite large. This is explained by the fact that the fraction of very low $Q^2$-regions contribution to energy loss very slowly decreases with lepton energy. Note that for τ-lepton the influence of these corrections is significantly larger.

For comparison of the influence of various corrections, three different models widely used for inelastic cross section calculations and estimations were chosen: Borog-Petrukhin formula [8], models of Bezrukov-Bugaev [9] and Bugaev-Shlepin [10] and also ALLM fit [11]. In Fig. 4a, the influence of the all above corrections for different models is presented. In Fig. 4b results for muon relative energy loss in standard rock calculated in a conventional way (the first scheme) for these models are compared with results of model [4] taking into account all corrections 2–5.

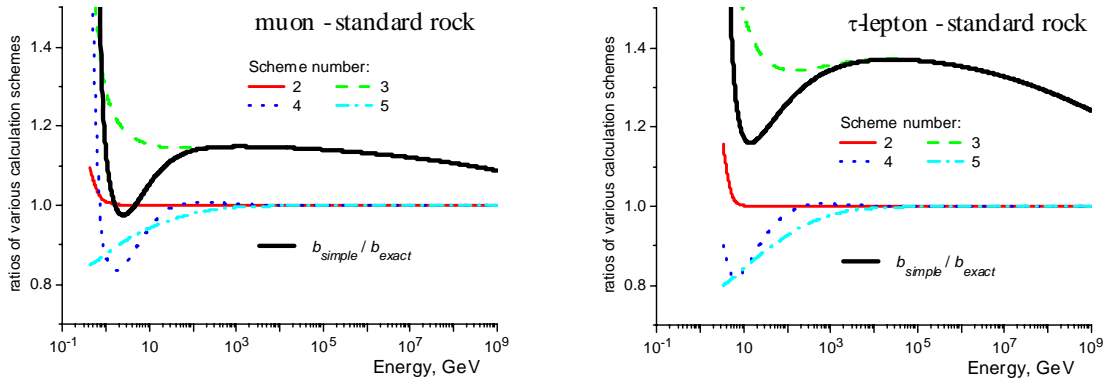

**Figure 3.** Ratios of $b_{in}$ for various schemes to exact one and a cumulative effect of all corrections (thick line).

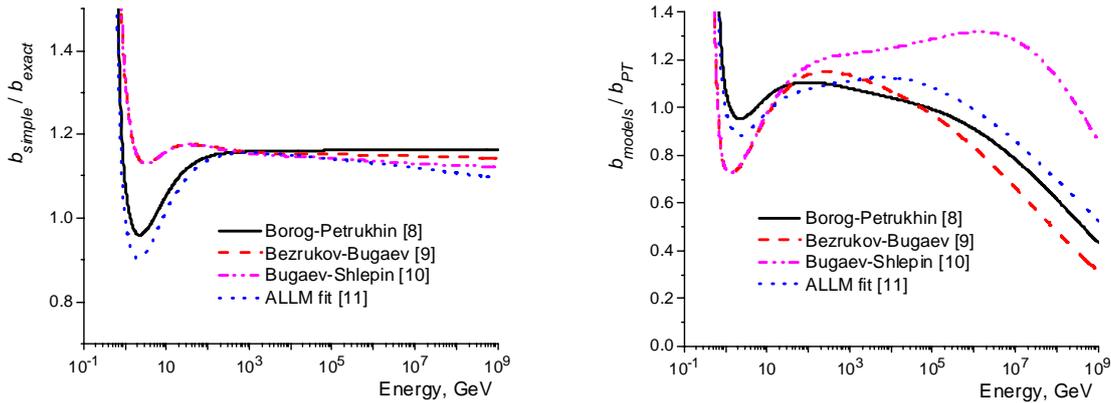

**Figure 4a.** Corrections for various models. For [9] and [10] models, nuclear effects are taken into account according to [9].

**Figure 4b.** Ratios of $b_{in}$ for various models. For $b_{model}$ the first (conventional) scheme was used.

As Fig. 4 shows, allowance for all corrections corresponding to items 2–5 (Section 1) leads to close results for different models of proton structure functions. This effect is comparable to and even exceeds the differences due to the choice of structure function model. Thus, conventional approximate approach to calculations of energy loss for heavy leptons in rock leads to significant distortions (up to tens percent) over a broad energy range.

## References

[1]  A. Okada et. al., Fort. Phys., **32**, 135 (1984).
[2]  B. Badelek, J. Kwiecinski, Rev. Mod. Phys., **68**, 445 (1996).
[3]  A.M. Cooper-Sarkar et al., Int. J. Mod. Phys. A, **13**, 3385 (1998).
[4]  A.A. Petrukhin, D.A. Timashkov, Phys. At. Nucl., **67**, 2216 (2004).
[5]  C.G. Callan, D.J. Gross, Phys. Rev. Lett., **22**, 156 (1969).
[6]  A.A. Petrukhin, D.A. Timashkov, Phys. At. Nucl., **66**, 195 (2003).
[7]  J. Gomez et al., Phys. Rev. D, **49**, 4348 (1994).
[8]  V.V. Borog, A.A. Petrukhin, Proc. XIV ICRC, Munich, 6, 1949 (1975).
[9]  L.B. Bezrukov, E.V. Bugaev, Sov. J. Nucl. Phys., **33**, 635 (1981).
[10]  E.V. Bugaev, Yu.V. Shlepin, Phys. Rev. D **67**, 034027 (2003).
[11] H. Abramowicz, A. Levy, hep-ph/9712415.